# Temperature dependent bilayer ferromagnetism in $Sr_3Ru_2O_7$

M. B. Stone, M. D. Lumsden, R. Jin, B. C. Sales, D. Mandrus and S. E. Nagler
*Oak Ridge National Laboratory, P. O. Box 2008, Oak Ridge, Tennessee 37831*
Y. Qiu
*National Institute for Standards and Technology, Gaithersburg, Maryland 20899*

The Ruthenium based perovskites exhibit a wide variety of interesting collective phenomena related to magnetism originating from the Ru $4d$ electrons. Much remains unknown concerning the nature of magnetic fluctuations and excitations in these systems. We present results of detailed inelastic neutron scattering measurements of $Sr_3Ru_2O_7$ as a function of temperature, probing the ferromagnetic fluctuations of the bilayer structure. A magnetic response is clearly visible for a range of temperatures, $T$ = 3.8 K up to $T$ = 100 K, and for energy transfers between $\hbar\omega$ = 2 and 14 meV. These measurements indicate that the ferromagnetic fluctuations manifest in the bilayer structure factor persist to surprisingly large temperatures. This behavior may be related to the proximity of the system in zero magnetic field to the metamagnetic/ferromagnetic transition.

I. INTRODUCTION

The collective magnetic and electronic behavior of the 4*d* transition metal oxides remains a forefront problem in condensed matter physics. The most widely studied systems to date are the Ruddlesden-Popper series of ruthenates, that are structural analogues of the various high-$T_C$ cuprates. A naïve comparison of the ruthenates to the cuprates might suggest that covalency and itinerant effects would be more dominant in the ruthenates, and these would tend to suppress "interesting" physical behavior typically associated with magnetism and proximity to metal insulator transitions. In practice, however, the physical behavior of the ruthenates is rich and diverse. The parent compounds $SrRuO_3$[1], $Sr_2RuO_4$[18,2,3], $Sr_3Ru_2O_7$ and $Sr_4Ru_3O_{10}$[4] as well as a series of doped derivative compounds[5,6,7,8,9] have been investigated and shown to offer a range of magnetic and electronic properties. These include metamagnetic transitions and quantum critical phenomena[10] and a putative exotic p-wave superconducting transition[18].

This paper focuses on magnetic fluctuations in the bilayer ruthenate compound, $Sr_3Ru_2O_7$. A metamagnetic phase transition has been observed in this system at low temperatures in moderate applied magnetic fields[10], and is described as being relevant to a quantum critical end point in the field-temperature phase diagram[11]. Although quantum critical points reside explicitly at $T = 0$, their influence can persist to finite temperatures, large compared to the energy scale of the relevant interactions[12,13]. In the case of $Sr_3Ru_2O_7$, measurements reveal a first-order like metamagnetic phase transition up to $T = 1.1$ K in an applied magnetic field[14,15], and further influence of these transitions on macroscopic properties such as transport and magnetization continues to higher temperatures, $T \approx 5$ K[10]. Other measurements, such as $^{17}O$-NMR[16] and Hall-effect transport[19], indicate that ferromagnetic behavior is present up to even higher temperatures, $T \approx 100$ K. Here we describe measurements using a spectral probe coupled directly to the magnetic excitations over a wide range of temperatures. We investigate the magnetic fluctuations associated with the bilayer structure of $Sr_3Ru_2O_7$ in regions that are considerably distant from the metamagnetic quantum critical point (QCP), $5 \leq T \leq 100$ K, yet we find that the ferromagnetic correlations within the bilayer persist to anomalously high temperatures compared to their energy scale. This behavior is likely linked to the proximity of $Sr_3Ru_2O_7$ to a ferromagnetic phase transition and the metamagnetic QCP.

$Sr_3Ru_2O_7$ is the $n = 2$ member of the Ruddlesden-Popper cubic perovskite series, $Sr_{n+1}Ru_nO_{3n+1}$.[17] The value of $n$ is the number of sequentially coupled Ru-O layers in the structural unit, ranging from the $n = 1$ member of the series, the $p$-wave superconductor $Sr_2RuO_4$[18], through the $n = \infty$ member, the $T_c = 160$ K ferromagnet, $SrRuO_3$. $Sr_3Ru_2O_7$ is a bilayer system as shown schematically in Fig. 1(a). The structure of these materials consists of layers of Ru-O and Sr-O alternating along the $c$-axis. The $4d$ $Ru^{4+}$ ions are in a corner-sharing octahedral coordination with surrounding oxygen atoms.

Based upon magnetic field and temperature dependent electrical transport measurements $Sr_3Ru_2O_7$ can be described as a paramagnetic quasi-two-dimensional metal[13,19] near a ferromagnetic transition. Pressure induced[32,29,20] magnetic transitions also exist in $Sr_3Ru_2O_7$, including a ferromagnetic phase associated with moderate uniaxial compression along the $c$-axis. Early thermodynamic measurements were controversial due to differences in sample growth techniques and difficulties associated with sample purity.[21,31] The precise boundaries of the metamagnetic transition[10] are dependent upon the direction of the applied magnetic field.[22] A diagram of the different physical regimes as determined from transport measurements[10] is reproduced in Fig. 1(b) with the metamagnetic transition observed in the vicinity of $H$ = 5 T. Advances in crystal growth and sample purity[23] have produced samples with additional crossovers associated with the metamagnetic quantum critical point at low-temperatures[24]. There is great interest in understanding the spectrum of magnetic fluctuations and their relationship to the metamagnetic QCP. Ultimately, this calls for single crystal inelastic neutron scattering measurements as a function of temperature and field magnitude and orientation in this compound.

Prior neutron scattering investigations of $Sr_3Ru_2O_7$ have examined the crystal structure and explored portions of the low-energy excitation spectrum.[25,26,28] Neutron powder diffraction measurements reported in Ref. 26 observed no sign of a structural phase transition from room temperature to $T$ = 9 K, and no indication of magnetic order for temperatures as low as $T$ = 1.6 K. We have examined single crystal samples with elastic neutron scattering using the cold neutron disk chopper spectrometer (DCS) at NIST, and have found no indication of magnetic order in the ($h0\ell$) scattering plane down to $T$ = 0.08 K in zero magnetic field.

Inelastic neutron scattering measurements described in Ref. 25 surveyed several high symmetry directions of reciprocal space with an emphasis on energy transfers between one and four meV. A fluctuation spectrum with a bilayer structure factor along $c^*$ was observed for wavevectors (0.75 0 $\ell$). A series of constant-energy scans within the ($hk$0) plane indicated excitations associated with incommensurate wavevectors at positions (1±$\zeta$ 0 0) and (1 ±$\zeta$ 0) where $\zeta$ = 0.25 and 0.09. These are described as being influenced by the bilayer structure and the small rotation of the octahedrally coordinated Ru ions. The incommensurate magnetic fluctuations were shown to dampen and move to commensurate wave-vectors with increasing temperature.

This paper presents new inelastic neutron scattering measurements that are complementary to those of the earlier study. The dynamics within a localized region of reciprocal space are probed over a wider range of temperatures and energy transfers. The results confirm that the low energy antiferromagnetic incommensurate fluctuations dampen with increasing temperature. However, the new results show that a simple relaxor function does not provide an adequate description of the data. In addition there is the surprising result that the ferromagnetic bilayer nature of the fluctuations measured along (1-$\zeta$ 0 $\ell$) persists to higher than expected temperatures.

## II. EXPERIMENTAL METHODS

Single crystals of $Sr_3Ru_2O_7$ were grown using an NEC SC-M15HD image furnace. For feed-rod preparation, a mixture of $SrCO_3$ and Ru, in the molar ratio of 4:3, was heated in flowing oxygen initially to 1175 C at a rate of 125 C/hr. The temperature was then increased to 1400 C at 25 C/hr. After remaining at 1400 C for 5 hours, the mixture was quenched to room temperature. The resulting material was reground and the powder was pressed into rods and heated in air at 1100 C for 12 hr. Single crystals were grown using a feed rate of 20 mm/hr. and a growth rate of 10 mm/hr. in an atmosphere of 10% oxygen and 90% argon.

We describe the crystal structure using a tetragonal unit cell consistent with space group *I4/mmm*[26,27,28]. In this description, there are four Ru ions per unit cell, and no distortion/rotation of the Ru-O octahedra. The measured low-temperature, ($T \leq 4$ K) lattice constants are $a = b = 3.87(1)$ Å and $c = 20.77(5)$ Å. As illustrated in Fig. 1(a), neighboring bilayer groups are offset by ½$a$ such that Ru ions in neighboring bilayers are not aligned with one another. The Ru-Ru separation is $a$ within the *ab* plan; the intra-bilayer Ru-Ru distance is ≈ 0.2$c$, and the inter-bilayer (bilayer center to bilayer center) distance is ½$c$. In Sr$_3$Ru$_2$O$_7$, the Ru-O octahedra are rotated away from the orthogonal crystalline *a*- and *b*-axes about the *c*-axis by approximately +/- 8°.[26,29,28]

Single crystal magnetization measurements were performed on a sample from the identical growth that produced our neutron scattering crystal to look for potential inclusions of other magnetic phases. Fig. 2 shows the temperature dependent magnetization in the presence of an external field of $H = 1$ Tesla applied perpendicular to the *c*-axis, i.e. magnetic field vector within the *ab*-plane. The high temperature portion (150 K < $T$ < 300 K) of these data can be fitted to a Curie-Weiss magnetic susceptibility with a temperature independent background with $\Theta_{CW} = -126(1)$K[30] and effective moment 2.50(5) $\mu_B$/Ru. These parameters and the location of the rounded maximum at $T \approx 17$ K are consistent with previous results[31,32]. Differences in the fitted Curie-Weiss temperature are potentially due to the sensitivity to different ranges of temperature used in the comparison to the model.[33]

Inelastic neutron scattering measurements were performed using the HB1 triple axis spectrometer at the High Flux Isotope Reactor at the Oak Ridge National Laboratory. The sample consisted of a roughly 4 gram single crystal of Sr$_3$Ru$_2$O$_7$ oriented in the (*h*0ℓ) scattering plane and mounted in a closed-cycle He$^4$ refrigerator. Horizontal collimation was chosen as 48'-60'-60'-240' between source and monochromator, monochromator and sample, sample and analyzer, and analyzer and detector respectively. The spectrometer was operated with fixed final energy, $E_f$ = 14.7 meV, using a pyrolytic graphite (PG 002) monochromator and analyzer. Pyrolytic graphite filters were placed after the sample to substantially reduce higher order spurious scattering processes. In this configuration the energy resolution at the elastic position, as measured from the incoherent scattering, is FWHM 1.5(1) meV. All measurements were made for fixed incident neutron monitor count.

Figure 3 shows a pair of typical constant-energy scans illustrating background corrections that are made to much of the data presented. The panels on the left show raw data, with the total background, $BG$, illustrated as a dotted line. The panels on the right have the smooth background subtracted, and the solid lines are guides-to-the-eye. The non-magnetic background is assumed to be the sum of a fast neutron background independent of wave-vector and energy transfer, an energy dependent constant term, and a scattering angle dependent term. The fast neutron background was determined as 1.6(2) counts per minute from measurements with the analyzer turned several degrees from the Bragg scattering condition. The energy dependent constant was determined by fitting as discussed below, and was found to be a monotonically decreasing function of energy and increasing function of temperature varying from 45(1) [57(2)] counts per 250 MCU[34] at 2 meV to 27(3) [34(1)] counts per 250 MCU at 14 meV at $T$ = 4 K [$T$ = 100 K]. The scattering angle dependent term is significant when the scattering angle approaches small values, below 12 degrees. This was quantified by measurements as a function of scattering angle for several different energy transfers for regions in reciprocal space with negligible magnetic scattering. For constant energy scans, the scattering angle is minimized when wave-vector transfer is minimized resulting in extra scattering intensity near $\ell = 0$ for scans along (0.75, 0, $\ell$). These background contributions have been subtracted from much of the data presented in this paper as noted in figure captions.

## III. RESULTS AND DISCUSSION

The present work includes a detailed investigation of the bilayer structure factor as a function of energy, wave-vector transfer, and temperature. In particular constant energy scans between (0.75 0 -8) and (0.75 0 +8) were performed for energy transfers between $\hbar\omega$ = 2 and 14 meV for $T$ = 3.8 K, 20 K, 50 K and 100 K. Figure 4 summarizes the results, showing that the data for $T$ = 3.8 K, 20 K and 50 K are qualitatively similar. As discussed below, these portions of the spectrum are well described by a ferromagnetic bilayer in the crystallographic *ab* plane. The scattering intensity is significantly diminished for the $T$ = 100 K data (Fig. 4(d)), however, the low-energy scattering intensity in the vicinity of $\hbar\omega \approx 2$ meV remains present. We note that there is significant phonon scattering at larger values of energy and wave-vector transfer. Additional scans were performed along the (0.95 0 $\ell$) direction for selected values of energy transfer. For $\hbar\omega \geq$ 2 meV, where contamination scattering from nearby nuclear Bragg peaks is unimportant, a similar bilayer structure factor is observed compared to the (0.75 0 $\ell$) measurements as shown in Fig. 5 for scans at 6 meV.

Measurements were also performed as a function of $h$ at the positions ($h$ 0 0) and ($h$ 0 ±4.8), corresponding to the central ($\ell = 0$) and first maxima ($\ell \approx 4.8$) of the bilayer structure factor as seen in Fig 5. Accounting for differences in instrumental resolution, the scans along ($h$ 0 0) are consistent with those reported by Capogna *et al.*[25] At 2 meV, we measure a single broad peak centered at (1 0 0), and at 4 meV we observe peaks at incommensurate positions $h = 1 \pm 0.18(2)$ with relative intensities consistent with the Ru magnetic form factor.

Measurements along ($h$ 0 ±4.8), Fig. 6, indicate a broad diffuse peak centered at $h = 1$ for energy transfers below 4 meV. This peak is much broader than the FWHM wave-vector resolution: $\delta h(\mathbf{Q} = (1\ 0\ -4.8), \hbar\omega = 4\ \text{meV}) = 0.068$ rlu. At 6 meV, the scattering intensity peaks at incommensurate positions symmetric about $h = 1$, $h = 1 \pm 0.22(1)$. Although $|\ell| = 4.8$, the small value of the $c^*$ reciprocal lattice parameter is such that the dependence of the intensity on $|Q|$ is consistent with the broad peaks being magnetic in origin.

Figures 7-10 are representative constant-energy scans at the (0.75 0 $\ell$) positions. Again, the region of higher scattering intensity is localized around wave-vectors (0.75 0 0) and (0.75 0 +/-4.5) at energy transfers below $\hbar\omega \approx 12$ meV. We compare our measurement to a two-dimensional magnetic bilayer model with ferromagnetic correlations between the individual planes making up the two layers,

$$I = A\,|F(Q)|^2 \cos^2(2\pi \ell z/c) + PH + BG, \quad (1)$$

where $A$ represents the overall amplitude of the magnetic bilayer scattering, $F(Q)$ is the Ru magnetic form factor[35], $c$ is the measured lattice constant, $2z$ is the distance between the bilayers, $PH$ models the observed phonon scattering apparent at large $\ell$, and $BG$ is the background discussed earlier with examples shown in Fig. 3.[25] We note that the only background fitting parameter is the energy dependent constant discussed earlier. The Ru magnetic form factor falls off rapidly as a function of $Q$, such that for constant-energy scans along (0.75 0 $l$), $|F(Q)|^2$ at $\ell \approx 8$ is approximately 20% of its peak value. This is an initial indication that the scattering intensity observed at larger values of $\ell$ (i.e. Figs. 4 and 7-10 for $\hbar\omega = 14$ meV and $|\ell| > 8$) is non-magnetic.

The increase in scattering at large values of $|\ell|$ can be understood as arising from phonons. To verify this, constant wave-vector scans were performed at (0.75 0 8.5) for different temperatures. These are illustrated in Fig. 11(a). Both the $T = 20$ K and $T = 100$ K data indicate potentially two peaks, one at $\hbar\omega \approx 7$ meV and a second at $\hbar\omega \approx 15$ meV. The high temperature $T = 200$ K measurement indicates an overall increase in spectral weight at lower energy transfers which, together with the large $|Q|$ of the observed scattering, is consistent with phonon scattering. Figure 11(b) shows the generalized susceptibility, $\chi''(Q, \hbar\omega)$, obtained by normalizing the data shown in Fig. 11(a) by the Bose occupation factor,

$$n(\hbar\omega) + 1 = \frac{1}{1 - \exp(-\hbar\omega/k_B T)}. \quad (2)$$

The resulting normalized temperature dependent data, shown in Fig. 11(b), collapses onto a single curve providing clear evidence of the phononic nature of the observed scattering. Recent Raman scattering measurements[36] have identified a $\approx$14.5 meV mode associated with $RuO_6$ octahedral rotation in $Sr_3Ru_2O_7$; this is the likely origin of the optical phonon we have observed in our measurements.

The phonon scattering in Figs. 7-10. is modeled assuming Gaussian lineshapes. Our model of the scattering intensity thus includes two free parameters describing the bilayer model ($A$ and $z$), a constant background, and a pair of Gaussian peaks at larger values of wave-vector transfer. As indicated by the lineshapes in Figs. 7-10, this model is able to accurately represent our measured excitations. The average value of the bilayer separation for all temperatures was determined to be $2z = 4.2(2)$ Å which agrees well with the calculated value, $2z = 4.056$ Å, based upon the $T = 4$ K measured $c$ lattice constant and prior crystal structure determination.[4] We observe no systematic variation in $2z$ as a function of temperature or energy transfer. This model of the characteristic bilayer structure factor fits the data well for all scans measured at $T \leq 50$ K. Even at $T = 100$ K, the data for $\hbar\omega = 2$ meV continues to be well described by this model. For slightly larger values of energy transfer at $T = 100$ K the magnetic spectral weight is instead shifted to a central peak at (0.75 0 0) that is well described by a Lorentzian lineshape. At and above $\hbar\omega = 10$ meV, there is no visible peak at $\ell = 0$.

The measurements at energy transfers up to 6 meV exhibit the surprising feature that a well-defined ferromagnetic bilayer structure factor persists for temperatures up to 50K. At energy scales comparable to that of thermal energies one would expect that fluctuations associated with the ferromagnetism would be substantially dampened. We note that other physical quantities measured in $Sr_3Ru_2O_7$ also exhibit hints of ferromagnetic behavior at such temperatures. Both Hall effect transport measurements[19] and $^{17}$O-NMR[37] indicate that ferromagnetic fluctuations are present in the temperature range in which we observe the low-energy ferromagnetic bilayer structure factor. It is interesting to consider whether this may be related in some way to the metamagnetic QCP, and we will return to this issue later.

It is illuminating to examine the dynamic susceptibility, $\chi''(\mathbf{Q}, \hbar\omega)$. Figure 12 shows a plot of $\chi''(\mathbf{Q}, \hbar\omega)$ at $T = 4$K, for the wave-vector (0.75,0,0) corresponding to the peak of the bilayer structure factor. These data were obtained by combining background subtracted constant energy scans (e.g., as shown in Fig. 7) and constant wave-vector scans normalized to a fixed monitor. The data from the constant energy scans is averaged over 5 points around the peak to improve the statistics. The signal obtained in this way is normalized by the Bose occupation factor (Eq. 2) to convert the result to a generalized susceptibility.

The imaginary susceptibility measured in several different ruthenate materials[38] has been found to be well-described using the single relaxor function,

$$\chi''(Q,\omega) = A \frac{\omega\gamma}{\omega^2 + \gamma^2}, \quad (3)$$

where $A$ is an arbitrary scale factor and $\gamma$ is a width parameter providing a characteristic energy scale. This expression has been applied to $Sr_3Ru_2O_7$ by Capogna *et al*[25]. The best fit of the $T = 4$ K data to Eq. 3 is shown in Fig. 12 as a dotted line. The fitted value of $\gamma = 2.1(3)$ meV agrees well that reported previously for $T = 1.5$ K.[25] We also compare our results to the response function of a damped harmonic oscillator (DHO),

$$\chi''(Q,\hbar\omega) = A \frac{\omega\Gamma}{(\omega^2 - \omega_0^2)^2 + (\omega\Gamma)^2}. \quad (4)$$

This function contains two distinct energy scales, a damping parameter, $\Gamma$, and $\omega_0$, a characteristic energy of the system.[39] The best fit of the $T = 4$ K data to Eq. 4 yields $\Gamma = 10(1)$ meV, $\omega_0 = 5.5(2)$ meV and is shown in Fig. 12 as a solid line. It is clear by inspection of Fig. 12 that the DHO function fits the data better than the single relaxor model, and this is supported by a comparison of the normalized chi-squared, $\chi^2$, for the fits which is smaller by a factor of 2.5 for the DHO function.

The data from other temperatures bears out this conclusion. We have also carried out constrained simultaneous global fits to the data from all temperatures. Here the parameters $A$ (Eq. 3 and 4) and $\omega_0$ (Eq. 4) are taken as common for all $T$, and the parameters $\Gamma$ (Eq. 4) and $\gamma$ (Eq. 3) are assumed to be linear functions of temperature. The resulting fits yield a normalized $\chi^2$ of 6.4 for the relaxor model, and 2.4 for the DHO. The global fit to the DHO model yields a characteristic energy $\omega_0 = 5.6(4)$ meV. The meaning of this energy scale is not entirely clear, but we speculate that it may be associated with the intra-bilayer coupling. It may be worth noting that the c-axis lattice constant and the bilayer thickness are particularly sensitive to temperature, first increasing as temperature decreases below $T = 300$ K then decreasing as the temperature is reduced below $T \approx 25$ K[29]. This lattice constant is directly related to the bilayer separation and therefore to the interaction between Ru moments in the bilayer. Additional theoretical developments would help to elucidate the issue.

## IV. CONCLUSIONS

We provide detailed inelastic neutron scattering measurements of the ferromagnetic fluctuations associated with the bilayer structure in $Sr_3Ru_2O_7$. The results show that a well defined bilayer structure factor unexpectedly persists up to temperatures of 50K. We also find that a DHO response function provides a better description of the imaginary part of the generalized susceptibility than does the relaxor model that has been previously applied to this system. It should be noted that the characteristic energy scale of the DHO, roughly 60 K, may provide a simple way to understand the persistence of the bilayer response. A well defined bilayer structure factor presumably will be washed out by thermal fluctuations only when $k_B T$ exceeds $\omega_0$.

There may be more general reasons why ferromagnetic fluctuations persist to unexpectedly high temperatures in $Sr_3Ru_2O_7$. Recent theoretical[12] and experimental[13] work on systems near QCP behavior show that under some circumstances, in particular when more than one natural energy scale is present, fluctuations usually associated with a $T = 0$ phase transition may be observable at high temperatures. One might ask whether the QCP associated with the metamagnetic transition and its proximity to ferromagnetism plays a role here. The consequences of $Sr_3Ru_2O_7$ being a *nearly ferromagnetic* paramagnet are subtle, and structural and lattice effects may also be very important. We hope that the results presented here will help stimulate further experimental and theoretical investigations leading to a more complete understanding of this surprisingly interesting system.

Acknowledgements: Oak Ridge National Laboratory is managed by UT-Battelle, LLC, for the U. S. Department of Energy under Contract No. DE-AC05-00OR22725. This work utilized facilities supported in part by the National Science Foundation under Agreement No. DMR-0454672.

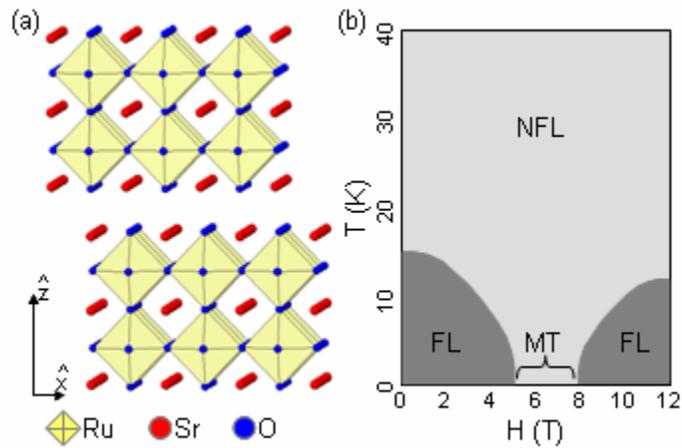

Figure 1. (Color online) (a) $Sr_3Ru_2O_7$ bilayer structure viewed along the *b*-axis, vertical direction indicates the *c*-axis. Octahedral coordination of the ruthenium ions with surrounding oxygen atoms is also indicated. (b) Temperature versus magnetic field behavior of $Sr_3Ru_2O_7$ as measured by Perry *et al.* (Ref. 10) via electrical transport with applied magnetic field parallel to the electrical current direction in the *ab* plane. NFL represents the disordered paramagnetic regime (non-Fermi liquid), FL corresponds to Fermi liquid behavior, and MT indicates a range of magnetic fields associated with the metamagnetic transition at low-temperatures. Figure does not represent a phase diagram, rather regions of different behavior.

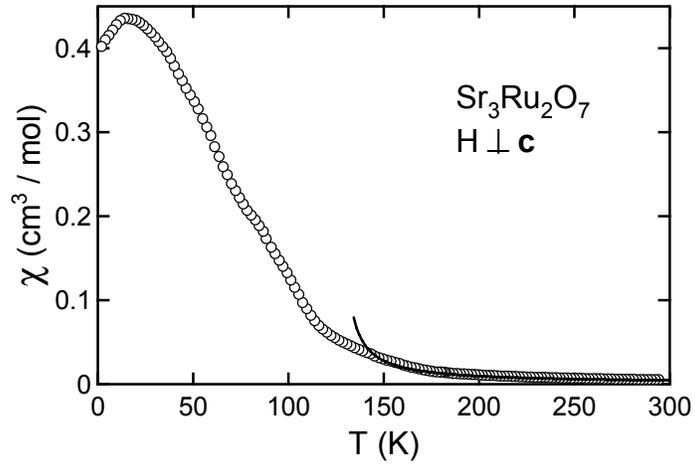

Figure 2. Magnetization as a function of temperature for $Sr_3Ru_2O_7$, measured with $H = 1$ T applied within the *ab* plane. Data have been plotted as $\chi \equiv M/H$. Solid line between $T = 150$ K and $T = 300$ K is a comparison to a Curie-Weiss high temperature magnetic susceptibility as discussed in the text. Curie-Weiss calculation is extended below $T = 150$ K for comparison.

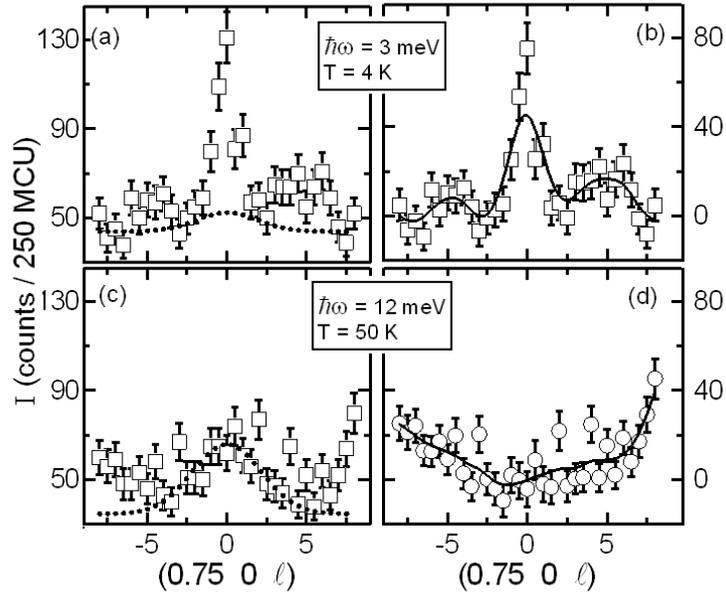

Figure 3. Comparison of two representative scans illustrating background subtraction. Panels on the left are raw data, with background indicated by dotted line. Panels on the right are the background subtracted data and the solid lines are guides-to-the-eye. Upper panels: $T = 4$ K, $\hbar\omega = 5$ meV. Lower panels: $T = 50$ K, $\hbar\omega = 12$ meV. The data have been scaled to a fixed monitor count of 250 monitor count units (MCU), corresponding to a counting time of approximately 4.1 (3.8) minutes per point at 3 meV (12 meV).

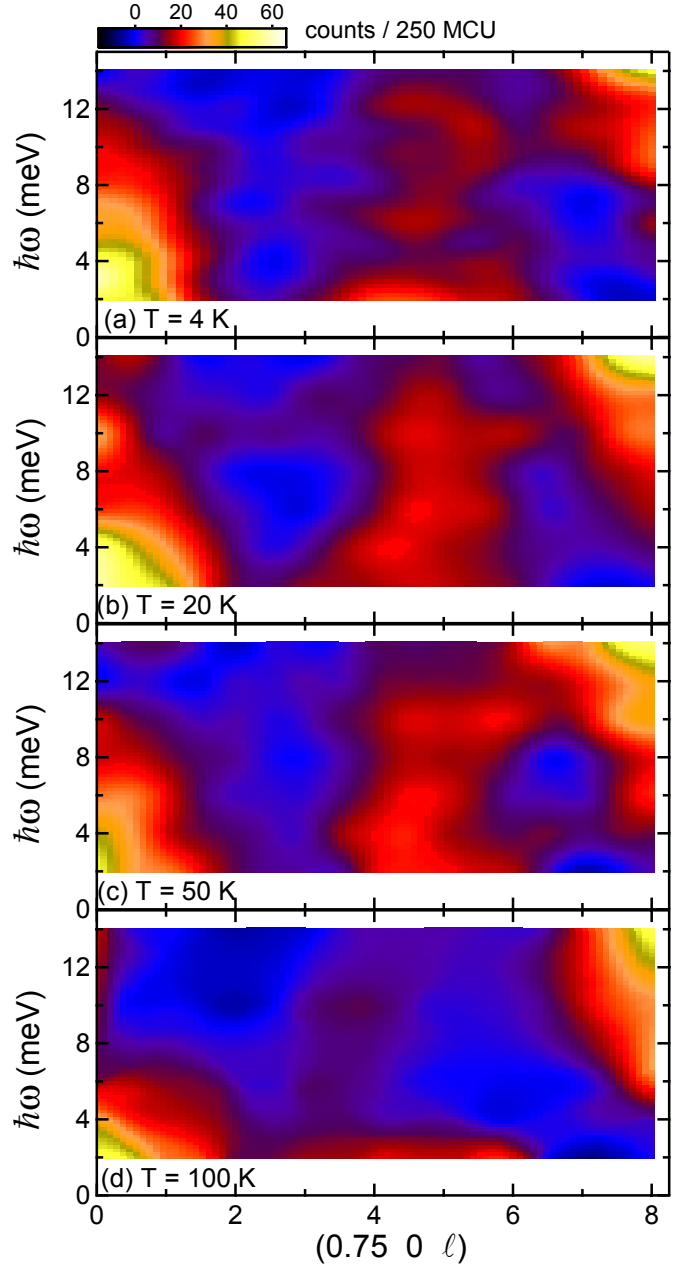

Figure 4. (Color) Background subtracted scattering intensity of $Sr_3Ru_2O_7$ as a function of energy transfer and wave-vector transfer along the $(0.75\ 0\ \ell)$ direction. Results are compiled by interpolating between measured values and using Gaussian smoothing for clarity of presentation. Data are averaged for $\pm\ell$. All measurements are plotted on the identical intensity scale. These can be compared with the representative full-range unaveraged constant-energy scans shown in Figs. 7-10. (a) $T = 4$ K. (b) $T = 20$ K. (c) $T = 50$ K. (d) $T = 100$ K.

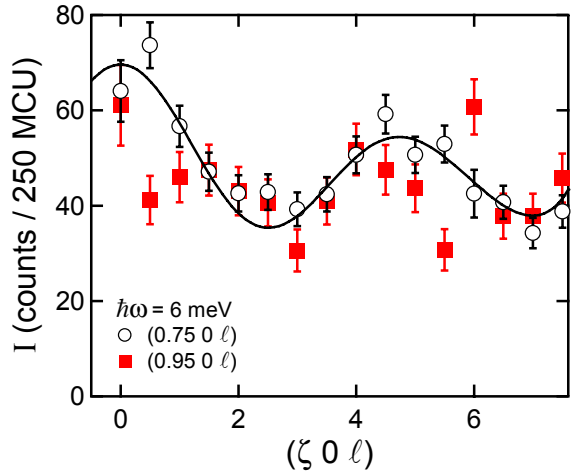

Figure 5. (Color online) $T = 4$ K scattering intensity of $Sr_3Ru_2O_7$ along the $(\zeta\ 0\ l)$ wave-vector at $\zeta = 0.75$ and $0.95$ for an energy transfer of 6 meV. Background subtraction consists of fast neutron background and sample $2\theta$ dependent background. Data were measured between $\ell = 8$ and $-8$ wave-vector transfer and are shown averaged for values of $+\ell$ and $-\ell$. Solid line is fit of a bilayer model to the data of the $(0.75\ 0\ \ell)$ scan, as discussed in the text.

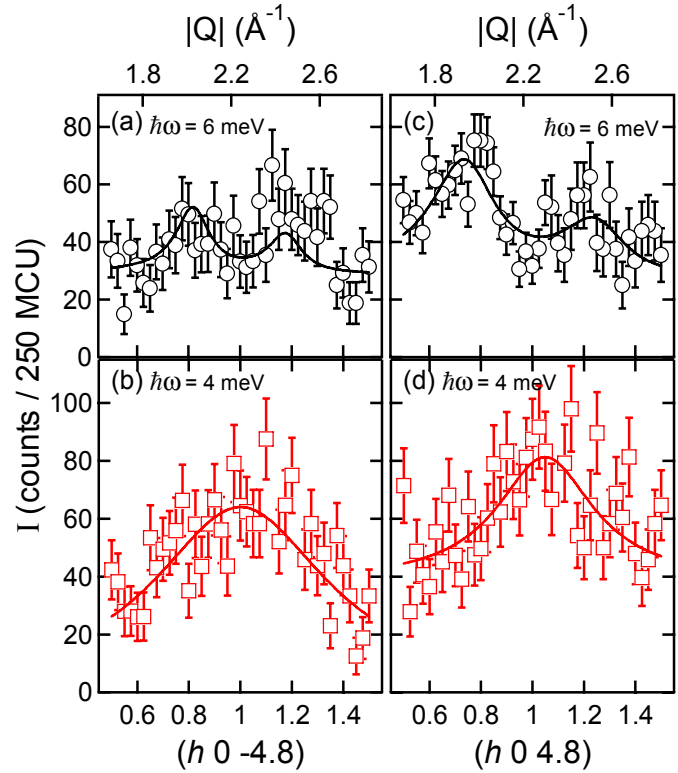

Figure 6. (Color online) $T = 4$ K scattering intensity of $Sr_3Ru_2O_7$ along the $(h\ 0\ \pm4.8)$ wave-vector at energy transfers of 4 and 6 meV. Background has been subtracted (see text). Solid lines are fits to a single Lorentzian ($\hbar\omega = 4$ meV) or a pair of symmetric Lorentizians ($\hbar\omega = 6$ meV) multiplied by the Ruthenium magnetic form factor and serve as guides to the eye.

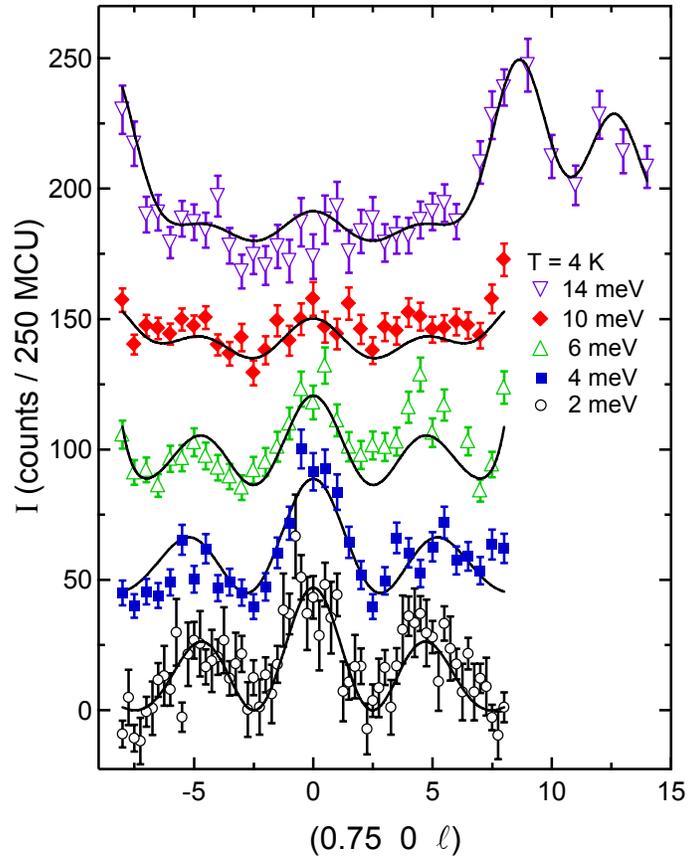

Figure 7. (Color online) $T = 4$ K constant-energy scans along $(0.75\ 0\ \ell)$. Data have been background subtracted and the ordinate has been offset in increments of 45 counts. As described in the text, solid lines are fits to the magnetic bilayer model with symmetric Gaussians located at $\ell \approx +/-\ 8.7$ and $+/-\ 13$.

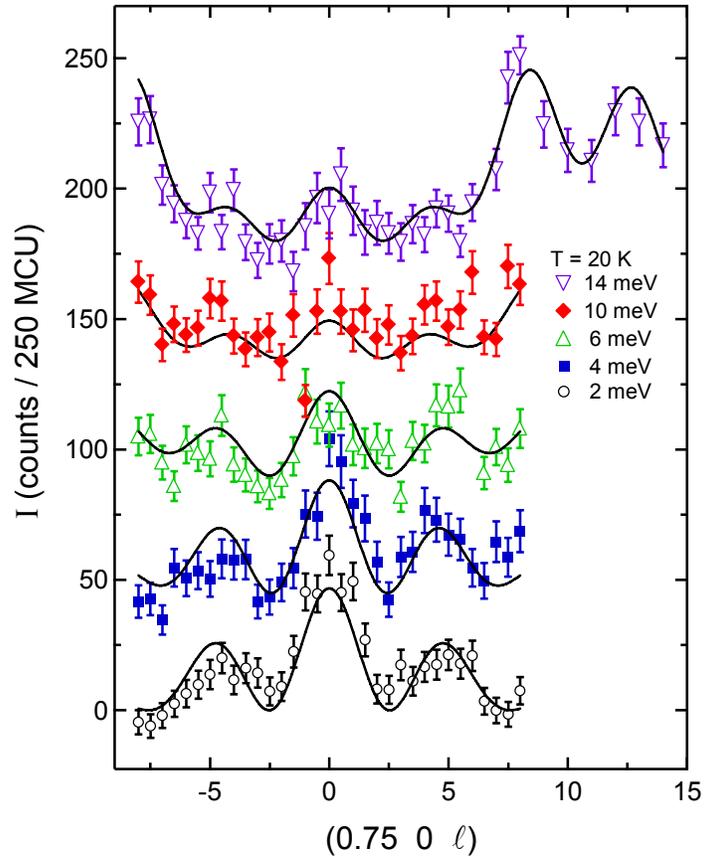

Figure 8. (Color online) $T = 20$ K constant-energy scans along $(0.75\ 0\ \ell)$. See Fig. 7 caption for additional information.

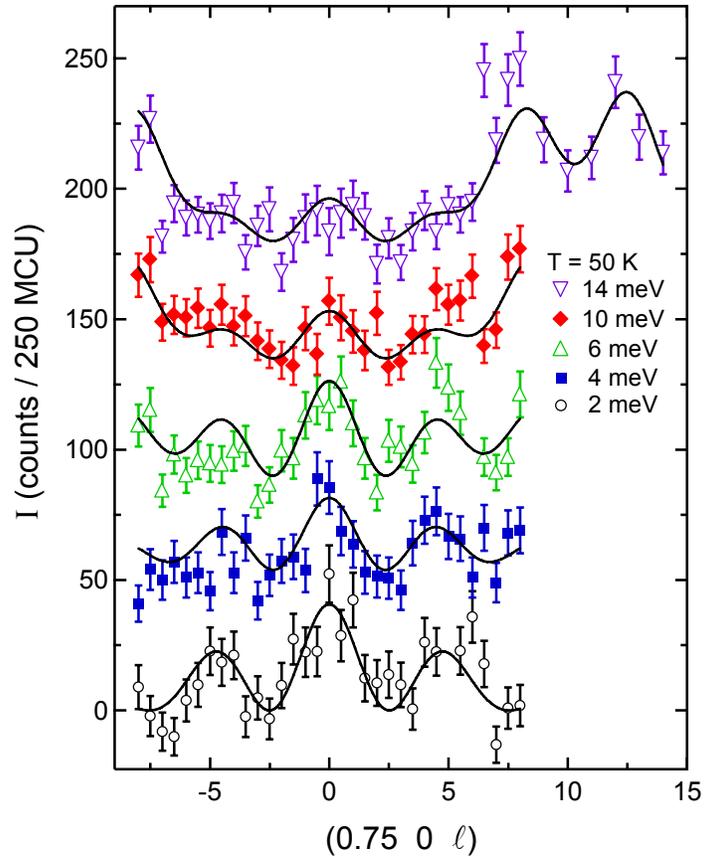

Figure 9. (Color online) $T$ = 50 K constant-energy scans along (0.75 0 $\ell$). See Fig. 7 caption for additional information.

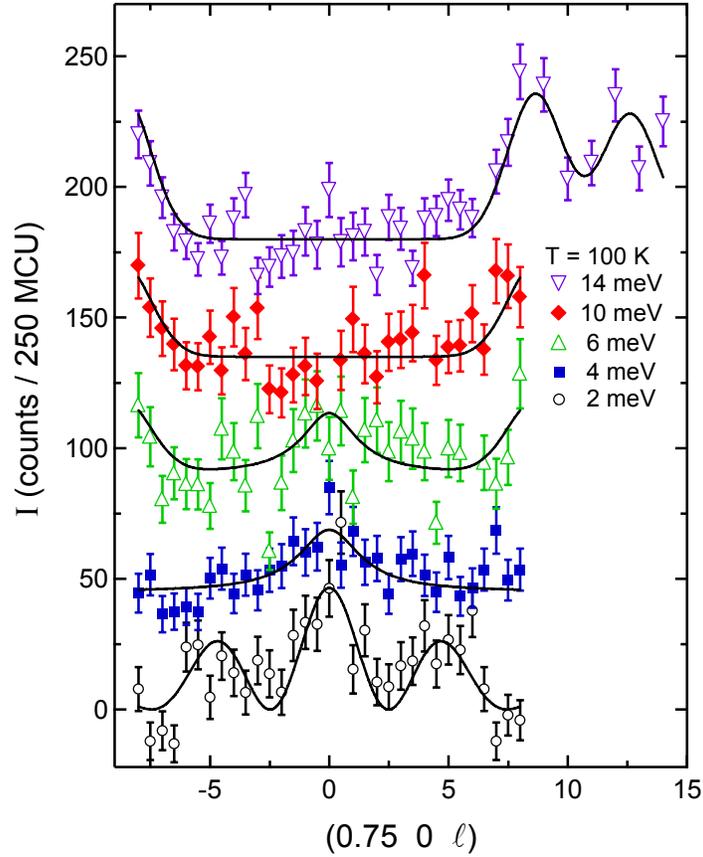

Figure 10. (Color online) $T$ = 100 K constant-energy scans along (0.75 0 $\ell$). Data have been background subtracted and ordinate offset in increments of 45 counts. Solid line in $\hbar\omega$ = 2 meV scan is a fit to the magnetic bilayer model described in the text. $\hbar\omega$ = 4 and 6 meV measurements are fit to a Lorentzian centered at (0.75 0 0). $\hbar\omega$ = 6, 10, and 14 meV includes symmetric Gaussians located at $\ell \approx$ +/- 8.7 and +/- 13.

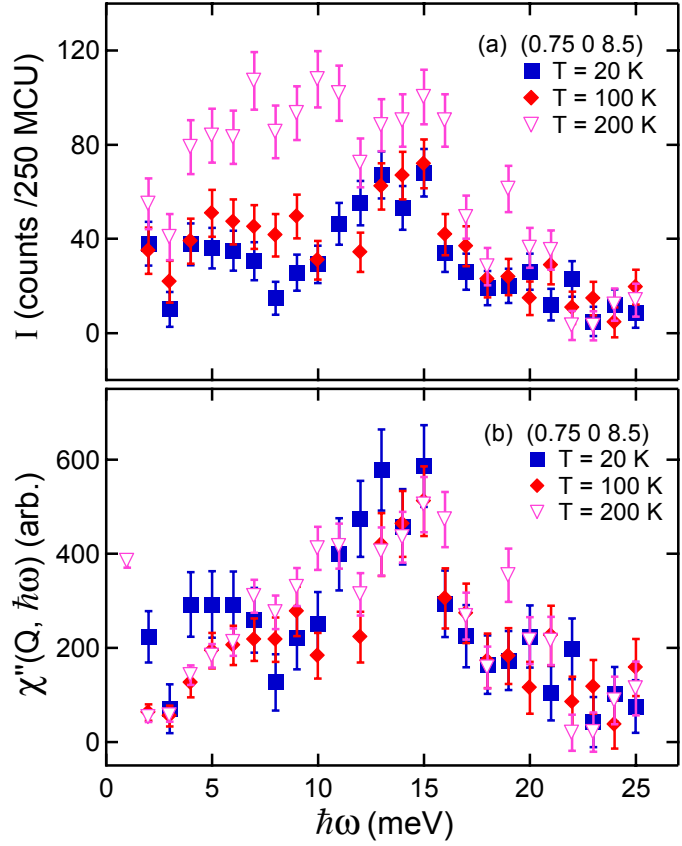

Figure 11. (Color online) (a) Background subtracted constant wave-vector scans illustrating phonon scattering at (0.75 0 8.5) at $T$ = 20 K, 100 K and 200 K. (b) Data in panel (a) normalized by $n(\hbar\omega)+1$ as described in the text.

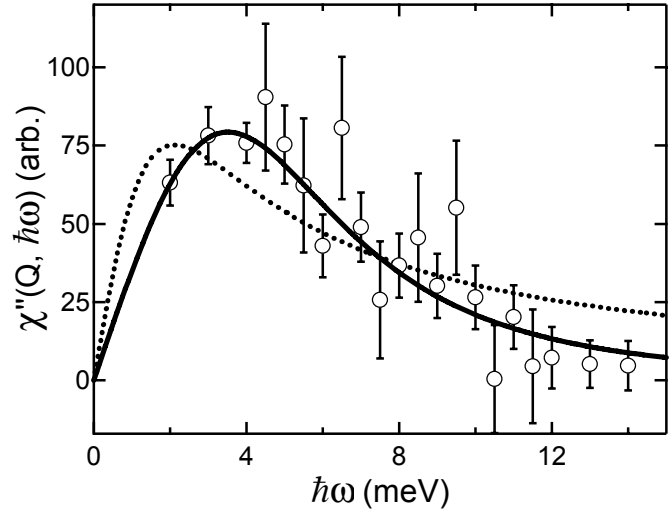

Figure 12. Generalized susceptibility, $\chi''(\mathbf{Q}, \hbar\omega)$, for $\mathbf{Q} = (0.75, 0, 0)$ and $T = 4$ K. Data points correspond to combining data from constant-energy scans as shown in Figs. 4 and 7 with constant wave-vector scans performed at (0.75, 0, 0). The solid line is a fit to the DHO response function, Eq. 4, and dotted line is a fit to the relaxor function, Eq. 3.

[1] A. P. Mackenzie, J. W. Reiner, A. W. Tyler, L. M. Galvin, S. R. Julian, M. R. Beasley, T. H. Geballe and A. Kapitulnik, Phys. Rev. B **58**, R13318 (1998). P. B. Allen, H. Berger, O. Chauvet, L. Forro, T. Jarlborg, A. Junod, B. Revaz and G. Santi, Phys. Rev. B **53**, 4393 (1996).

[2] A. P. Mackenzie, R. K. W. Haselwimmer, A. W. Tyler, G. G. Lonzarich, Y. Mori, S. Nishizaki and Y. Maeno, Phys. Rev. Lett. **80**, 161 (1998).

[3] T. Oguchi, Phys. Rev. B **51**, R1385 (1995); D. J. Singh, Phys. Rev. B **52**, 1358 (1995); M. Braden, W. Reichardt, S. Nishizaki, Y. Mori and Y. Maeno, Phys. Rev. B **57**, 1236 (1998).

[4] M. K. Crawford, R. L. Harlow, W. Marshall, Z. Li, G. Cao, R. L. Lindstrom, Q. Huang and J. W. Lynn, Phys. Rev. B **65**, 214412 (2002) and G. Cao, L. Balicas, W. H. Song, Y. P. Sun, Y. Xin, V. A. Bondarenko, J. W. Brill, S. Parkin and X. N. Lin, Phys. Rev. B **68**, 174409 (2003).

[5] S. Nakatsuji and Y. Maeno, Phys. Rev. Lett. **84**, 2666 (2000). S. –C. Wang, H. –B. Yang, A. K. P. Sekharan, S. Souma, H. Matsui, T. Sato, T. Takahashi, C. Lu, J. Zhang, R. Jin, D. Mandrus, E. W. Plummer, Z. Wang and H. Ding, Phys. Rev. Lett. **93**, 177007 (2004).

[6] A. Mamchik and I.Wei Chen, Phys. Rev. B **70**, 104409 (2004) and A. Mamchik, W. Dmowski, T. Egami and I.Wei Chen, Phys. Rev. B **70**, 104410 (2004).

[7] Y. Yoshida, S-I. Ikeda, H. Matsuhata, N. Shirakawa, C. H. Lee, S. Katano, Phys. Rev. B **72**, 054412 (2005).

[8] S. Ikeda, Y. Maeno, T. Fujita, Phys. Rev. B **57**, 978 (1998).

[9] A. Kanbayasi, J. Phys. Soc. Jpn. **44**, 108 (1978), and T. Kiyama, K. Yoshimura, K. Kosugl, H. Michor and G. Hilscher, J. Phys. Soc. Jpn. **67**, 307 (1998).

[10] R. S. Perry, L. M. Galvin, S. A. Grigera, L. Capogna, A. J. Schofield, A. P. Mackenzie, M. Chiao, S. R. Julian, S. Ikeda, S. Nakatsuji, Y. Maeno and C. Pfleiderer, Phys. Rev. Lett. **86**, 2661 (2001), and S. A. Grigera, R. S. Perry, A. J. Schofield, M. Chiao, S. R. Julia, G. G. Lonzarich, S. I. Ikeda, Y. Maeno, A. J. Millis and A. P. Mackenzie, Science **294**, 329 (2001).

[11] A. J. Millis, A. J. Schofield, G. G. Lonzarich, and S. A. Grigera, Phys. Rev. Lett. **88**, 217204 (2002).

[12] A. Kopp and S. Chakravarty, Nature Physics **1**, 53 (2005).

[13] B. Lake, D. A. Tennant, C. D. Frost and S.E. Nagler, Nature Materials, **4**, 329 (2005).

[14] S. A. Grigera, P. Gegenwart, R. A. Borzi, F. Weickert, A. J. Schofield, R. S. Perry, T. Tayama, T. Sakakibara, Y. Maeno, A. G. Green and A. P. Mackenzie, Science **306**, 1154 (2004).

[15] M. Chiao, C. Pfleiderer, R. Daou, A. McCollam, S. R. Julian, G. G. Lonzarich, R. S. Perry, A. P. Mackenzie and Y. Maeno Cond-mat/0207697.

[16] K. Kitagawa, K. Ishida, R. S. Perry, T. Tayama, T. Sakakibara and Y. Maeno, Phys. Rev. Lett. **95**, 127001 (2005).

[17] S. N. Ruddelsden and P. Popper, Acta Crystallogr. **10**, 538 (1957); *ibid*, **11**, 54 (1958).

[18] Y. Maeno, H. Hashimoto, K. Yoshida, S. Nishizaki, T. Fujita, J. G. Bednorz and F. Lichtenbeerg, Nature (London) **372**, 532 (1994).

[19] Y. Liu, R. Jin, Z. Q. Mao, K. D. Nelson, M. K. Haas and R. J. Cava, Phys. Rev. B **63**, 174435 (2001).

[20] S. I. Ikeda et al., Physica C **364-365**, 376 (2001); M. Chiao et al. Physica B **312-313**, 698 (2002); Y. V. Sushko, B. DeHarak, G. Cao, G. Shaw, D. K. Powell and J. W. Brill, Solid State Comm. **130**, 341 (2004); S-I. Ikeda, N. Shirakwaw, T. Yanagisawa, Y. Yoshida, S. Koikegami, S. Koike, M. Kosaka, and Y. Uwatoko, J. Phys. Soc. Jpn. **73**, 1322 (2004). H. Yaguchi, R. S. Perry, and Y. Maeno. Cond-mat/0510383.

[21] M. Itoh, M. Shikano, and T. Shimura, PRB **51**, 16432 (1995); G. Cao, S. McCall, and J. E. Crow. Phys. Rev. B **55**, R672 (1997); S. I. Ikeda and Y. Maeno Physica B **259-261**, 947 (1999).

[22] S. A. Grigera, R. A. Borzi, A. P. Mackenzie, S. R. Julian, R. S. Perry and Y. Maeno, Phys. Rev. B. **67**, 214427 (2003).

[23] I. Ikeda, J. Cryst. Growth **237-239**, 787 (2002); R. S. Perry and Y. Maeno, J. Crystal Growth **271**, 134 (2004).

[24] E. Ohmichi, Y. Yoshida, S. I. Ikeda, N. V. Mushunikov, T. Goto, and T. Osada, Phys. Rev. B **67**, 024432 (2003); Z. X. Zhou, S. McCall, C. S. Alexander, J. E. Crow, P. Schlottmann, A. Bianchi, C. Capan, R. Movshovich, K. H. Kim, M. Jaime, N. Harrison, M. K. Haas, R. J. Cava, and G. Cao. Phys. Rev. B **69**, 140409(R) (2004); P. Gegenwart, F. Weickert, M. Garst, R. S. Perry and Y. Maeno, Cond-Mat/0507359.

[25] L. Capogna, E. M. Forgan, S. M. Hayden, A. Wildes, J. A. Duffy, A. P. Mackenzie, R. S. Perry, S. Ikeda, Y. Maeno and S. P. Brown, Phys. Rev. B. **67**, 012504 (2003).

[26] Q. Huang, J. W. Lynn, R. W. Erwin, J. Jarupatrakorn and R. J. Cava, Phys. Rev. B. **58**, 8515 (1998).

[27] Hk. Müller-Buschbaum and J. Wilkens, Z. Anorg. Allg. Chem. **591**, 161 (1990).

[28] H. Shaked, J. D. Jorgensen, O. Chmaissem, S. Ikeda and Y. Maeno, J. Solid State Chem. **154**, 361 (2000).

[29] H. Shaked, J. D. Jorgensen, S. Short, O. Chmaissem, S.-I. Ikeda, Y. Maeno, Phys. Rev. B **62**, 8725 (2000).

[30] errors in parentheses correspond to a single standard deviation.

[31] R. J. Cava, H. W. Zandbergen, J. J. Krajewski, W. F. Peck, Jr., B. Batlogg, S. Carter, R. M. Fleming, O. Zhou and L. W. Rupp, Jr., J. Solid State Chem. **116**, 141 (1995).

[32] S. I. Ikeda, Y. Maeno, S. Nakatsuji, M. Kosaka and Y. Uwatoko, Phys. Rev. B **62**, R6089 (2000).

[33] We note that, as mentioned in Ref. 32, the negative value of $\Theta_{CW}$ should not be considered as indicative of there being solely antiferromagnetic correlations in itinerant magnetic systems.

[34] MCU are monitor count units. 250 MCU corresponds to a counting time of approximately 3.6 minutes for energy transfers between 5 meV and 8 meV.

[35] In the absence of an *a priori* calculation of the $Ru^{4+}$ magnetic form factor, we approximate by using the magnetic form factor for the $Ru^+$ ion. Any differences associated with this are less than our experimental error.

[36] M. N. Iliev, V. N. Popov, A. P. Litvinchuk, M. V. Abrashev, J. Bäckström, Y. Y. Sun, R. L. Meng and C. W. Chu, cond-mat/0408432.

[37] K. Kitagawa, K. Ishida, R. S. Perry, T. Tayama, T. Sakakibara and Y. Maeno, Phys. Rev. Lett. **95**, 127001 (2005).

[38] Y. Sidis, M. Braden, P. Bourges, B. Hennion, S. NishiZaki, Y. Maeno and Y. Mori, Phys. Rev. Lett. **83**, 3320 (1999); M. Braden, Y. Sidis, P. Bourges, P. Pfeuty, J. Kulda, Z. Mao and Y. Maeno, Phys. Rev. B **66**, 064522 (2002); O. Friedt, P. Steffens, M. Braden, Y. Sidis, S. Nakatsuji and Y. Maeno, Phys. Rev. Lett. **93**, 147404 (2004).

[39] P. M. Chaikin and T. C. Lubensky *Principles of condensed matter physics*, Cambridge University Press, (1997).